# Elastically driven Kelvin-Helmholtz-like instability in planar channel flow

Narsing K. Jha[1] and Victor Steinberg[1,2*]

[1]Department of Physics of Complex Systems, Weizmann Institute of Science, Rehovot 76100, Israel

[2]The Racah Institute of Physics, Hebrew University of Jerusalem, Jerusalem 91904, Israel.

*Corresponding author. Email: <u>victor.steinberg@weizmann.ac.il</u>

**Abstract:**

Kelvin–Helmholtz instability (KHI) is widely spread in nature on scales from micrometer up to Galactic one. This instability refers to the growth of perturbation of an interface between two parallel streams of Newtonian fluids with different velocities and densities, destabilized by shear strain and stabilized by density stratification with the heavier fluid at the bottom. Here, we report the discovery of the purely elastic KH-like instability in planar straight channel flow of viscoelastic fluid, which is theoretically considered to be stable. However, despite the remarkable similarity to the Newtonian KHI temporal interface dynamics, the elastic KHI reveals qualitatively different instability mechanism. Indeed, the velocity difference across the interface strongly fluctuates and non-monotonically varies in time due to energy pumping by elastic waves, detected in the flow. A correlation of the elastic wave intensity and efficiency of the elastic KHI in different regimes suggests that the competition between the destabilizing factor of the elastic waves and the stabilizing effect of the elastic stress difference generated by the velocity difference at the interface is the novel instability mechanism of the elastic KHI.

**One Sentence Summary:** Elastic Kelvin-Helmholtz-like instability reveals similarity in temporal dynamics with Newtonian Kelvin-Helmholtz instability, but the novel instability mechanism drastically differs.

Originally, the Kelvin–Helmholtz instability (KHI) is referred to the perturbations of interface between two parallel streams of Newtonian fluids with different velocities and densities, and the heavier fluid is at the bottom. Being one of the most studied shear flow instabilities and widespread in natural systems like Saturn's bands, oceanographic, and meteorological flows *(1,2)* and even on the Galactic scales explaining an expanding ring of young stars, gas and dust *(3,4)*. It is generally destabilized by the shear strain and stabilized by stratification effect. This name is currently used to describe more general instabilities of free shear flows or mixing layer with continuous variations of velocity and density over a finite layer thickness.

The elastic stress in viscoelastic fluid flows generated by the polymer stretching and its role in KHI is explored for decades with the conclusion that no pure elastic instability occurs at Re<1 and the Weissenberg number Wi>>1, another control parameter in the problem defining the degree of polymer stretching *(5,6)*. Here, both Re=$\rho$UL/$\eta$ and Wi=$\lambda$U/L are defined via the mean fluid velocity U and the vessel size L, $\rho$, $\eta$ and $\lambda$ are the density, dynamic viscosity and longest polymer relaxation time of the fluid, respectively.

The first stability analysis of KHI in a free shear (mixing) layer of viscoelastic fluid at ***Re***>>1 and ***Wi***>>1 *(7)*, formed by two counter-propagating fluid streams, shows that the elastic stress has a stabilizing effect *(7)*, in agreement with the experiments in mixing layer flows *(8)*. Further, it has been shown that hoop stress generated by the interface perturbations may be replaced by an elastic membrane stabilizing free shear flow, similar to two-fluid Newtonian streams with surface tension at the interface (see appendix in *(7))*. However, more recent theoretical investigation of viscoelastic planar mixing layer *(9)* demonstrates, in contrast, the destabilizing effect of elastic stress at large elasticity ***El=Wi/Re***, while at moderate ***El≈1*** the flow is stabilized by elastic stress in accord with *(7)*.

We recently studied mixing layer elastic instability and transition to ET in viscoelastic fluid flow between two widely spaced cylinders in a channel flow at Re<<1 and Wi>>1 *(10)*. As the result of the instability, two mixing layers are formed by two large scale counter-rotating vortices with non-uniform shear velocity profiles containing two inflection points. It is suggested that the latter may lead to the mixing layer instability resulting in an intermittent appearance of small vortices and the growth of spatiotemporal averaged vorticity near the inflection points in elastic turbulence (ET) regime that maybe associated with a possibility of the elastic KHI *(10)*.

Two recent indications, numerical *(9)* and experimental *(10)*, of a possibility to observe purely elastic KHI in the mixing layer of viscoelastic fluid at Re<1 and Wi>>1 give a hope that more convincing and quantitative confirmation of the elastic KHI will resolve the issue. The first experimental observations of elastic instability for viscoelastic fluid flow strongly perturbed at the inlet were reported in *(11)* for pipe flow and in *(12-14)* for straight square microchannel. Above the transition large velocity fluctuations *(11,12)* further associated with an increase in flow resistance and a power-law decay in kinetic energy spectrum (13,14). These findings encourage to study the physical mechanism of instability in straight channel flows.

Here, we report strong quantitative evidence of the elastically driven KH-like instability in a planar straight channel flow of viscoelastic fluid in the ET regime at Re<<1 and Wi>>1. As the result of our recent investigation of the elastic instability and ET in straight channel viscoelastic fluid flow, the discovery of weakly unstable coherent structures (CSs) of coexisting stream-wise streaks and rolls self-organized into cycling process leading to stochastically steady state is reported (15). Here we show that an elastically driven instability, which destroys streaks in each cycle, is the elastically-driven KH-like instability that strongly recalls the Newtonian KHI temporal interface dynamics but with a qualitatively different instability mechanism.

## Results

We use a dilute polymer solution as the viscoelastic fluid in a straight planar channel flow of aspect ratio 7. Addition of the small amount of long flexible polymers into a fluid strongly affects the fluid flow at **Wi** >>1 and **Re**<<1, resulting in elastic instability and further ET *(6,16-18)*. Though instabilities and ET are observed in flows with curvilinear streamlines, they vanish in the limit of zero curvature *(6,19)*. ET is a chaotic, inertialess flow driven solely by nonlinear elastic stress generated by stretching of polymers due to the flow, which is strongly modified by a feedback reaction of the elastic stress *(18)*. A back reaction of the elastic stress field on the flow leads to a stationary stochastic state of ET.

In our recent study *(15)*, we report the discovery of self-organized coherent structures (CSs) due to purely elastic instability (Re<<1) in a planar straight viscoelastic channel flow, namely stream-wise streaks and stream-wise rolls. These coherent structures participate in self-sustained cycling process (SSP) and it is synchronized with elastic waves pumping energy to CSs. In the present study, we show that during each cycle, the streaks, a coherent structure of the stream-wise counter-propagating velocity fluctuations $u'$ in a frame moving with the mean velocity $u_m(z)$, appear first (primary instability), followed by the perturbations of the interface between two counter-propagating streaks. The interface perturbations lead to streak destruction by the instability, which is secondary instability and visually greatly recalls KHI widely known in Newtonian shear flows. This process repeats itself in a cycling manner locked with elastic waves period.

We now present the results of the analysis of the data obtained in the same experimental setup as used in *(15)* and its schematic is shown in Fig. 1**,** whereas measurement techniques, solution preparation and characterization of the polymer solution, the same as in *(15)*, are presented in Materials and Methods section. Experiments are conducted in a large aspect ratio channel flow of visco-elastic fluid and spatio-temporal flow structures are visualized to quantify the coherent structures (CSs) and instability mechanisms. Figure 2 demonstrates a series of snapshots of the streaks in span-wise plane taken at 16 moments for the same spatial window during a single cycle. We here concentrate on the temporal development of interface instability (Movie 1 (**Suppl. Mater.**)). In general, we also observed multiple streaks, but it is a more complicated case (Fig. S1 (**Suppl. Mater.**)) and is not the focus of present study. The cycle starts by slight interface perturbations of 20-30 µm initial height (in particular, in Fig. S2 (**Suppl. Mater.**)). Temporal dependence of amplitude and wavelength of quasi-periodic interface perturbations can be followed in Figs. 2, 3A, S2 (and in Fig. S3 in more details (**Suppl. Mater.**)). The amplitude is quantified in the plot in Fig. 3A with an initial linear increase, which changes to a short sharp rise leading to saturation and a consequent final steep increase towards the end of the cycle. The first rise is associated with the exponential growth, while the second strong one is linked to a billow creation in the mixing layer due to the generation of a vortex string, and further, a final rise is due to the complete streak destruction. The perturbation wavelength (Fig. S3 (**Suppl. Mater.**)) is constant in average but strongly fluctuating with time, $\Lambda$ =0.35±0.15 mm (Fig.S4A (**Suppl. Mater.**)). Furthermore, the velocity fluctuation profile across interface is not strictly sharp but rather continuous, so that a fit of the interface profile by u'/U ~ tanh z/$d$ provides $d$=0.1±0.02 mm at Wi=185 (Fig. 3B), where 2$d$ is the interface width or free shear layer thickness. Similar plots of the interfaces at Wi=55 and 325 are shown in Fig.S4B, C (**Suppl. Mater.**), and dependence of $d$ on Wi is shown in Fig. 3C, whereas the width of a resulting mixing or billow layer is about 0.1w=0.35 mm (Fig. S5 **Suppl. Mater.**).

To better illustrate the vorticity generation resulting from interface instability, the field of wall normal vorticity fluctuations ($\omega'$) are presented in Fig.4 for the same data as shown in Fig.2. The formation of a vortex string is clearly seen in the third row at z/w≈-0.1 in Fig. 4. The vortex string replaces the sharp interface shown in the first row in Fig. 2 that is beautifully observed during the cycle (Movie 2 (**Suppl. Mater.**)). However, to separate the vorticity by the rotating field from that of the vorticity by shear strain and also to improve the presentation of the vortices, we used squared of the imaginary eigenvalue of the velocity gradient tensor *(17)* that exclusively provides rotational part of the vorticity (Fig. 5C). Then, except the vortex string vorticity in the third image in Fig. 5C, a contribution of a rotational motion at the streak interface is seen in the first and second images. Another way to characterize interface instability is to examine temporal evolution of $\Delta u = u_4 - u_2$, the velocity difference at two specific locations across the interface (Fig. S6 (**Suppl. Mater.**)), shown for two consecutive cycles in ET at Wi=185 in Fig. 5A**.** Each cycle shows a non-monotonic temporal variation of $\Delta u$ during a cycle having the same peak values $\Delta u_{max}$ and cycle period and $\Delta u(t)$ varies surprisingly similar between first and second cycle even though the flow is highly unsteady. To estimate the value of a control parameter of interface instability, we consider the pure elastic KHI-like of a mixing layer at Wi=185 (Figs. 3B) with u'/U ~ tanh z/d and average shear strain $\Delta u_{max}/2d \approx 75$ s$^{-1}$ (d = 0.1mm; $\Delta u_{max} \approx 15$ mm/s at Wi=185, Fig 3C and 5A). As the result, one gets $Wi_{ml} = \lambda \Delta u_{max}/2d = 1275$ in a mixing layer, which is very large and might play a role of the elastic KHI-like control parameter.

We present the temporal dependence of $\Delta u(t)$ in three regimes: the transition at Wi=55, the drag reduction (DR) at Wi=325 together with ET at Wi =185 (Fig. 5B), which displays similar behaviour. There are different cycle periods at Wi=55, 185 and 325, in accord with corresponding frequencies of elastic waves *(15, 22)*, and about 5 times smaller values of $\Delta u_{max}$ for the transition compared with ET and DR regimes. It correlates with the dependence of the peak values in the energy spectrum of the elastic waves, which pump energy into ET and critical for the existence of CSs (Fig. 5D and E) (15, 22). However, in the DR regime at Wi=325, the peak value of $\Delta u_{max}$ is similar to ET but the elastic wave peak in the energy spectrum is about twice lower (Fig. 5D and E). Then significant interface perturbations in DR show up already at the beginning of a cycle, and a fast and complete break-down takes place without any KH-like billow formation (Fig. S7, S8 (**Suppl. Mater.**)). On the other hand, at Wi=55, where both peaks $\Delta u_{max}$ and of the elastic waves in the power spectrum are almost order of magnitude lower than in ET, the interface perturbations remain small for the whole cycle (Fig. S7, S8 (**Suppl. Mater.**)). These observations strongly support the suggestion about the critical role of the elastic waves in the destabilization of the interface.

**Discussion and Conclusion:**

We present above the solid evidence of the elastically driven KH-like instability. It is discovered in a planar straight channel flow of viscoelastic fluid at Re<<1 and Wi>>1, particularly in elastic turbulence (ET). The elastic KHI occurs in self-organized CSs, namely of stream-wise streaks, which participates in self-sustained cycling process and is synchronized by the energy pumping by elastic waves. During each cycle, the streaks appear first, and then it is followed by the perturbations of the interface between two counter-propagating streaks. The perturbations lead to its destruction by the instability, which visually strongly recalls the conventional KHI, though of purely elastic nature. Subsequently, after destruction of streaks, a new similar cycle repeats itself.

The temporal sequences of the snapshots during two cycles illustrate an interface perturbation growth until its destruction via generation of vortex string that recalls the conventional KHI. Perturbation growth starts with an exponential growth and then a sharp rise during the roll-up with a final rise due to the destruction of streaks. Despite a remarkable similarity in the temporal interface evolution in the purely elastic and inertial KHI, significant differences between these two types of instability are revealed. Firstly, in a contrast to the Newtonian KHI, where the velocity difference across the interface ($\Delta u$) is the only destabilized factor in the flow, in the elastic KHI, $\Delta u$ generates the elastic stress difference at the interface, which stabilizes the instability (7). Secondly, in the elastic KHI, $\Delta u$ strongly fluctuates and non-monotonically varies during a cycle. Thirdly and most importantly, $\Delta u_{max}$ along with elastic wave peak intensity determines the maximum energy pumping into the elastic KHI, which correlates with the destabilization of interface. Therefore, the results suggest that the interaction of the interface perturbations with elastic waves, detected in the flow, is the new key destabilizing factor in the elastic KHI that overcomes the stabilizing effect of the elastic hoop stress generated by interface perturbation curvature.

To summarize, we report the discovery of the purely elastic KH-like instability taking place in a planar straight channel flow of viscoelastic fluid at $Re \ll 1$ and $Wi \gg 1$. The temporal dynamics of the interface perturbations until its complete destruction remarkably recalls the conventional Newtonian KHI evolution. However, the velocity difference across the interface generates the elastic stress difference that stabilizes the flow, in a sharp contrast to the inertial KHI, where $\Delta u$ destabilizes it. Its peak value $\Delta u_{max}$ together with the suggested phase-locked energy pumping by the elastic waves and their interaction with vorticity of the perturbed interface are identified as the novel destabilizing factor that overcomes the stabilizing effect of the elastic hoop stress. The possible way of interaction of elastic wave and streaks for inertialess flow can be seen as the amplification of perturbations at shear layer interface by the simultaneous transverse oscillation and the stream-wise motion of the elastic wave. We are currently working on this instability mechanism. To conclude, despite the similarity to the inertial KHI temporal dynamics, the purely elastic KH-like instability demonstrates the completely different instability mechanism.

**Materials and Methods**

**Preparation and characterization of polymer solution**: As a working fluid, a dilute polymer solution of high molecular weight polyacrylamide (PAAm, Mw = 18 MDa; Polysciences) at concentration c = 80 ppm (c/c* ≃ 0.4, where c* = 200 ppm is the overlap concentration for the polymer used (22)) is prepared using a water-sucrose solvent with sucrose weight fraction of 64%. The solvent viscosity, $\eta_s$, at 20 °C is measured to be 130 mPa·s in a commercial rheometer (AR-1000; TA Instruments). Addition of polymer to the solvent increases the solution viscosity, $\eta$, of about 30%. The stress-relaxation method *(25)* is employed to obtain the longest relaxation time ($\lambda$) of the solution and it yields $\lambda = 17 \pm 0.5$ s.

**Set-up and Flow discharge:** Transparent straight channel of dimensions length (l)× Width (w)× Height (h) = $475 \times 3.5 \times 0.5$ $mm^3$ is made of PMMA and is sandwiched between two stainless steel cover plates. It had the array of cylindrical obstacles to perturb the incoming elastic stress field. It is organized in 3 stream-wise rows with 11 cylinders (N) in each row (span-wise) of 100μm diameter (D) and separated by 200μm (s) both directions and produced a strong perturbation (10-13, 22). Our straight channel has quasi-2D rectangular geometry, and its length ratio l/h=950 and is about 3 times larger and aspect ratio (w/h) is =7, which is 7 times larger than that of the square channel used in (10-12). As stated earlier, it is perturbed across the full channel width to uniformly excite the flow. The stream and span-wise coordinates of the channel are x and z, respectively, with (x, z) = (0, 0) lying at the center of the downstream cylinder and channel. The polymer solution is driven by the Nitrogen gas pressurized up to 100 psi.

The fluid exiting the channel outlet is weighed instantaneously m(t) as a function of time t by a PC-interfaced balance (BA210S, Sartorius) with a sampling rate of 5 Hz and a resolution of 0.1 mg. The time-averaged fluid discharge rate is <Q>=<Δm/Δt>. The Weissenberg and Reynolds numbers are defined as Wi=λU/h and Re= Uhρ/η, and U=<Q>/ρwh with fluid density ρ = 1310 Kg m−3.

**Imaging system and μ-time resolved PIV:** For flow visualization, the solution is seeded with red fluorescent particles of diameter 3 μm (R0300B, Thermoscientific) and illuminated uniformly by 2.5 W green laser (MGL-F-CNI Laser) at 532 nm wavelength. The region is imaged in the mid-plane via a microscope objective (Leitz Wetzlar 2.5X/0.008 and 4X/0.12), and a fast camera from Photron (mini UX 100) with a spatial resolution of 1.2 Mpxl at a rate of up to 3200 fps. Images are acquired with low and high spatial resolutions for temporal velocity power spectra and flow structure, respectively. Three hundred thousand velocity fields are used to obtain velocity power spectra. We perform particle image velocimetry (PIV) using PIVlab (23) to obtain the spatially resolved velocity field $\vec{u} = (u, w)$ in the several regions at different distances from the obstacle array. The acquired images were processed using FFT based correlation with four-step box size refinement, with the final box size being 32pxl × 32pxl (~64 × 64 µm2). Correlation box overlap is maintained at 50 % in all cases, with each box having roughly at least 4–6 particles to ensure strong correlations. We also used PIV to measure velocity power spectra and captured images with resolution 640X192 pixels. We used three hundred thousand velocity field to extract temporal velocity for capturing well-resolved spectra.

**Acknowledgements.** We thank T. Zaki for an inspiring discussion and Vijay Kumar for the help with the experiments. PBC Fellowship at Weizmann institute for NKJ is gratefully acknowledged. **Funding:** This work is partially supported by the grants from Israel Science Foundation (ISF; grant #882/15 and grant #784/19) and Binational USA-Israel Foundation (BSF; grant #2016145).. **Competing interests.** The authors declare no competing interests. **Data and materials availability:** All data in the main text or the supplementary materials is available on request.

## Supplementary Materials

Figures S1-S8,

Movies S1, S2

References (19-21)

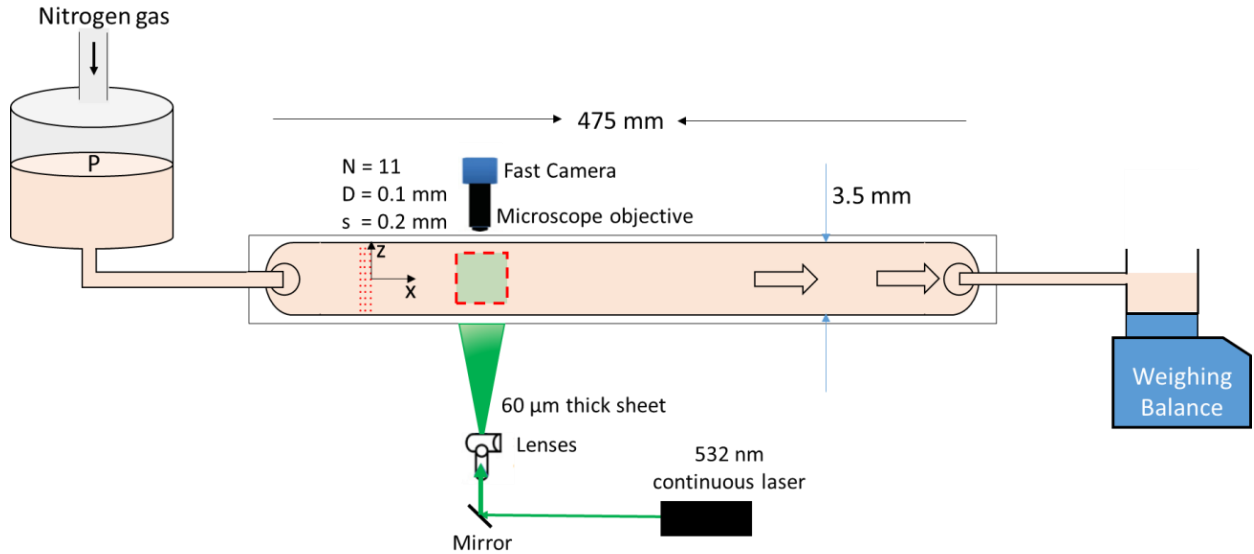

**Fig. 1. Schematic of experimental setup.** Straight channel of dimensions l × w × h = $475 \times 3.5 \times 0.5 \ mm^3$ made of PMMA has the array of cylindrical obstacles of 100μm diameter (D) organized in 3 rows with 11 cylinders (N) in each row separated by 200μm (s) both stream- and span-wise. Our straight channel has quasi-2D rectangular geometry, and its length ratio l/h=950 is about 3 times larger and aspect ratio w/h=7 is 7 times larger than that of the square channel used in (10-12). The polymer solution is driven by the Nitrogen gas pressurized up to 100 psi. To obtain the friction factor the fluid average velocity <U>=Q/ρwh is calculated via the average fluid discharge Q=<Δm/Δt>, where m(t) is weighed instantaneously as a function of time. Illuminated by a laser sheet, velocity field is measured via PIV.

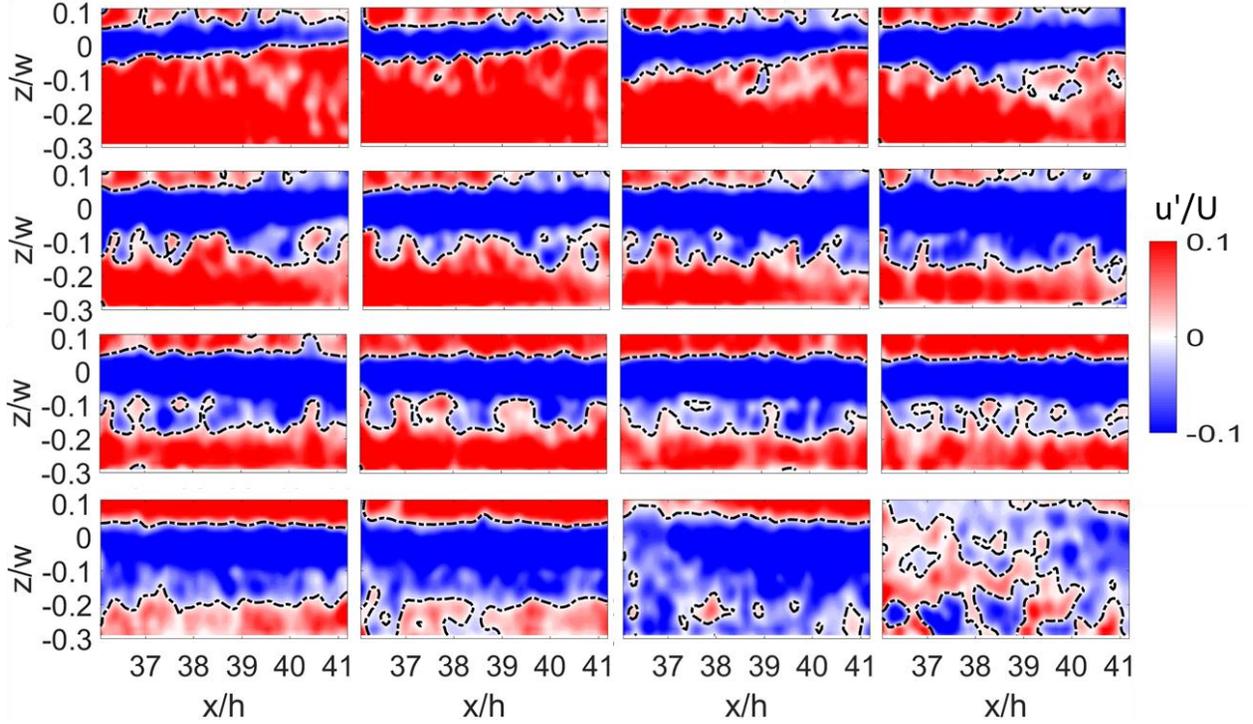

**Fig. 2. A cycle of the KH-like instability of the counter-propagating streaks interface in ET at Wi=185 and the location from the last row of obstacles normalized by the height x/h = 36-41.** Fluctuating stream-wise velocity field (u') in the frame moving with the mean velocity $u_m(z)$ normalized by the mean stream-wise flow velocity U (spatio-temporal mean velocity from weighing balance) is shown with their scales shown on the right at 16 moments of a single cycle at t* of: 0.15, 0.16, 0.19, 0.21 (first row left to right), 0.23, 0.24, 0.27, 0.30 (second row left to right), 0.33, 0.36, 0.44, 0.50 (third row left to right), 0.61, 0.64, 0.73 and 0.88 (fourth row left to right), where t* =t/$T_{cycle}$, $T_{cycle}$ = 225 msec. A coherent structure (CS) of stream-wise streaks of high and low speed are shown as positive and negative stream-wise fluctuating velocity (u') separated by dotted black lines of u'=0, presenting also span-wise perturbations, which are resulting from the Kelvin-Helmholtz-like perturbations. In the paper, we discuss the temporal development of the perturbations appeared at the lower interface between the flow against the mean stream (blue) and the flow in the direction of the mean stream (red) (also Movie 1 (**Suppl. Mater.**)).

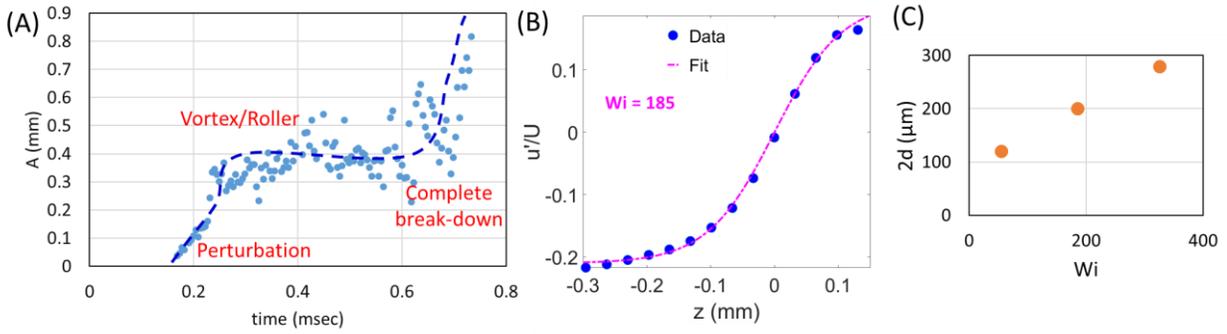

**Fig. 3. Characterization of the streak interface and the temporal and spatial development of its perturbations.** (A) shows a temporal growth of perturbation amplitude: initially, the amplitude increases linearly, after a short sharp rise, it saturates, and then it follows by a steep amplitude growth related to streak break-down. (B) shows the data and the fits of velocity profiles on the interface at Wi of 185. Finally, (C) presents the linear dependence of the shear layer thickness obtained by the fits on Wi.

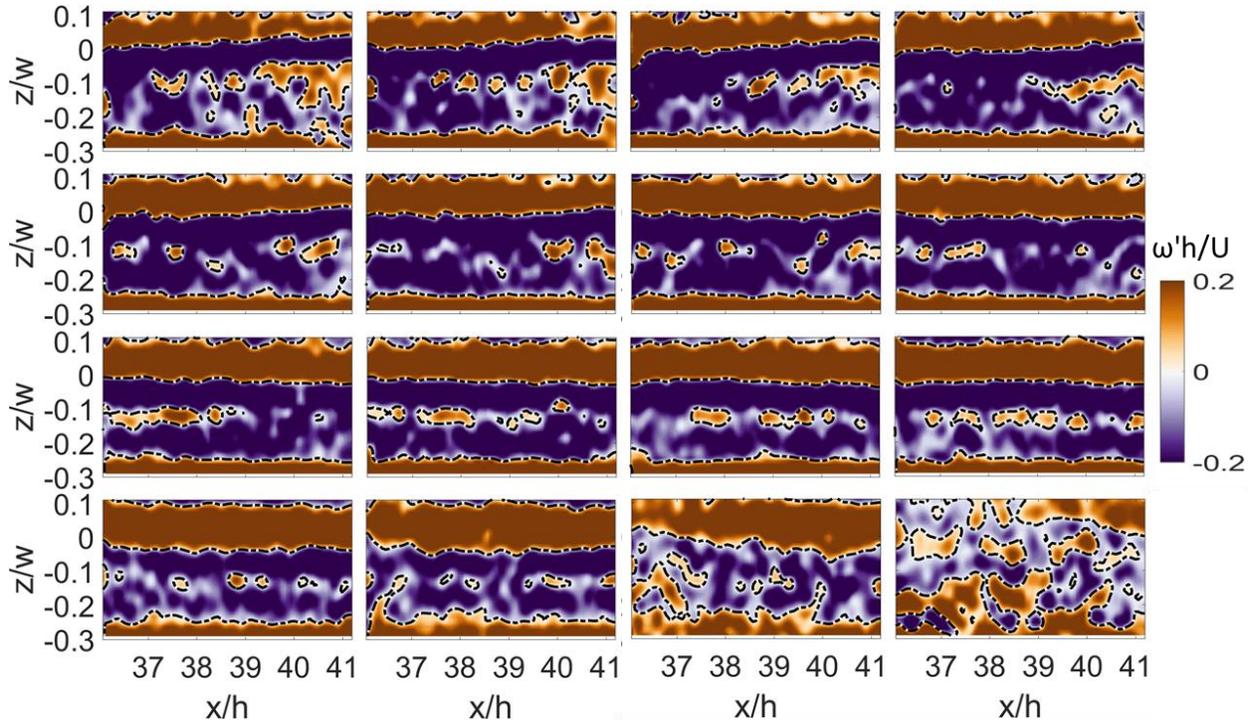

**Fig. 4**. **A cycle of the Kelvin-Helmholtz-like instability for the same data as in Fig. 1 presented by the field of the vertical vorticity fluctuations ω'.** In $\omega'$ plot, strips of vertical vorticity, generated by the elastically driven KH-like instability, at the center of streaks together with its random patterns near the streak edges and the vorticity of the shear strain in the bulk of steaks are shown. It is emphasized in the first snapshot, where a random pattern of $\omega'$ is marked by blue dotted line. The formation of a vortex string, clearly seen in the third row at z/w≈-0.1. The string of vortices replaces the sharp interface observed in the first row in Fig. 1, and it is also beautifully revealed during the cycle (also Movie 2 (**Suppl. Mater.**)).

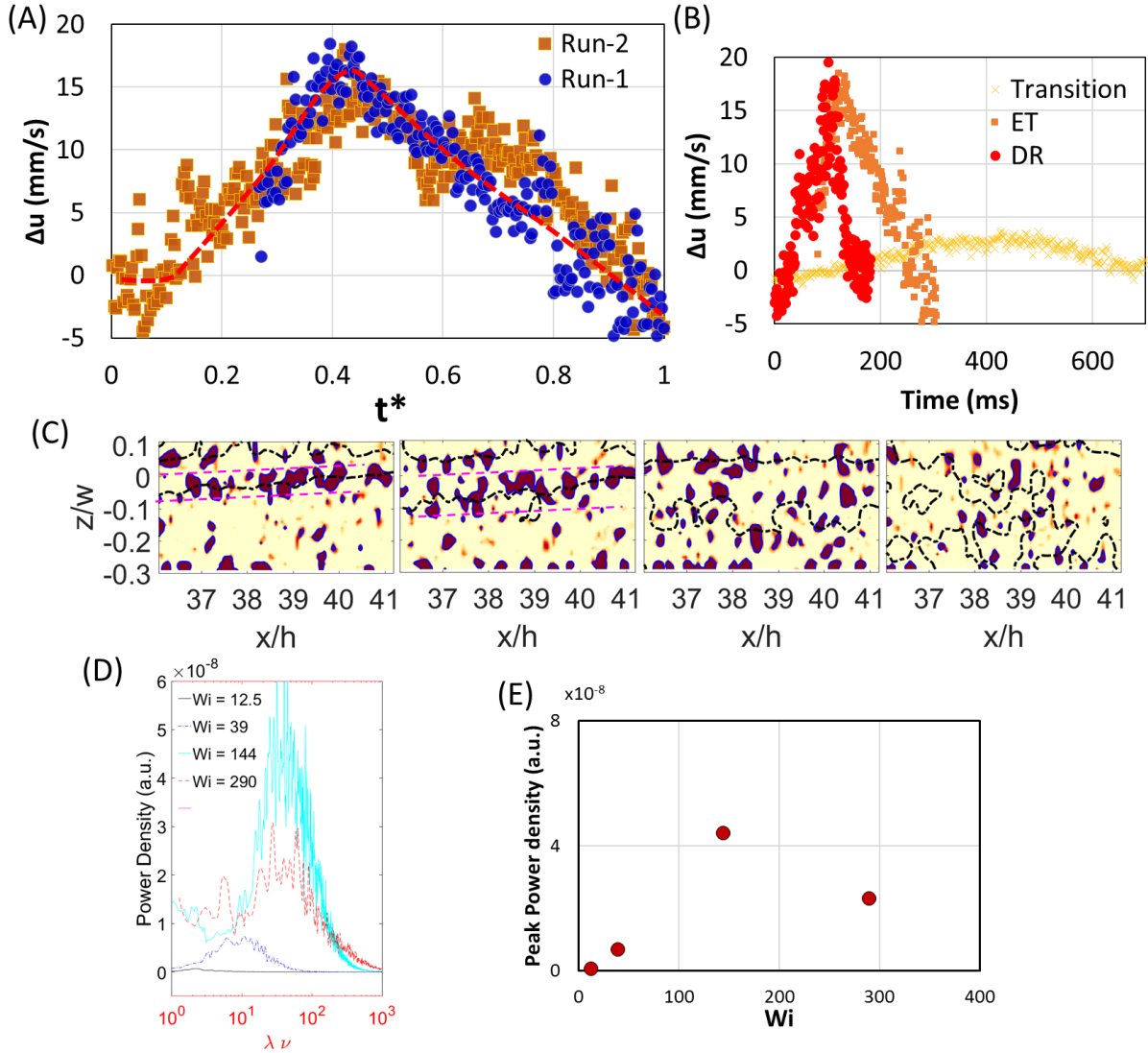

**Fig. 5. Further characterization of the streak interface temporal dynamics.** (A) Characterization of the elastic KHI by analyzing temporal evolution of Δu=u4-u2, where Δu is the velocity difference at two specific locations across the interface (Fig. S3 (**Suppl. Mater.**)), is shown for two consecutive cycles in ET at Wi=185. Their periods and the maximum values of Δu are close, also the shapes of Δu variations during the cycles are surprising similar after we shifted the time for Run-1 such that velocity peaks at the same time. (B) presents a comparison of Δu variation during a cycle in three different regimes: transitional at Wi=55, ET at Wi=185 and drag reduction (DR) at Wi=325. Their cycle periods $T_{cycle}$ differ significantly, since, as found in Ref.[15], $T_{cycle} \approx 1/f_{el}$, where $f_{el}$ is the frequency of the elastic waves that depends on Wi[15,22]. The maximum values of Δu are almost the same at Wi=185 and 325 but about 5 times higher than at Wi=55. Figure (C) presents the analysis of the fluctuating vorticity field to separate the vorticity by the rotating field from that of a shear strain by calculating the imaginary eigenvalue of the velocity gradient tensor. Then in such a way, we visualize the string of isolated vortices and vorticity of rotational flow near the curved streak edges and omit the vorticity related to the shear strain as shown in four snapshots at t* =0.15, 0.19, 0.36 and 0.88 (from left to right and then bottom). Only the snapshot at t*=0.36 should show the vortex string that indeed can be identified at z/w≈-0.1. (D) Span-wise velocity w power spectra in lin-log scale to compare the maximum values of energy peaks of elastic wave at different Wi in different flow regimes at l/h=38.6 are presented. (E) The plot of the peak values in the velocity power spectra vs Wi shows increase till ET and then decreases for DR regime.

# Supplementary Materials for:

## Elastically driven Kelvin-Helmholtz-like instability in planar channel flow

Narsing K. Jha[1] and Victor Steinberg[1,2*]

*Correspondence to: victor.steinberg@weizmann.ac.il

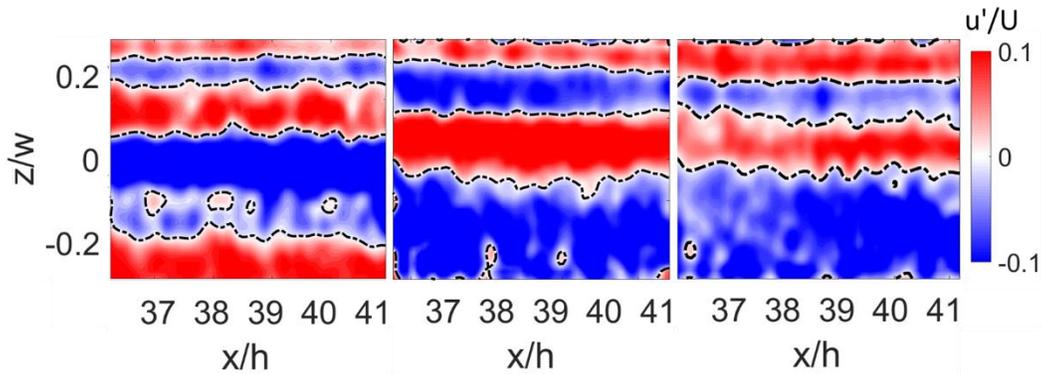

*Fig. S1. Examples of multiple streaks at different instants, which is a more complicated case of the elastic KHI with multiple variables.*

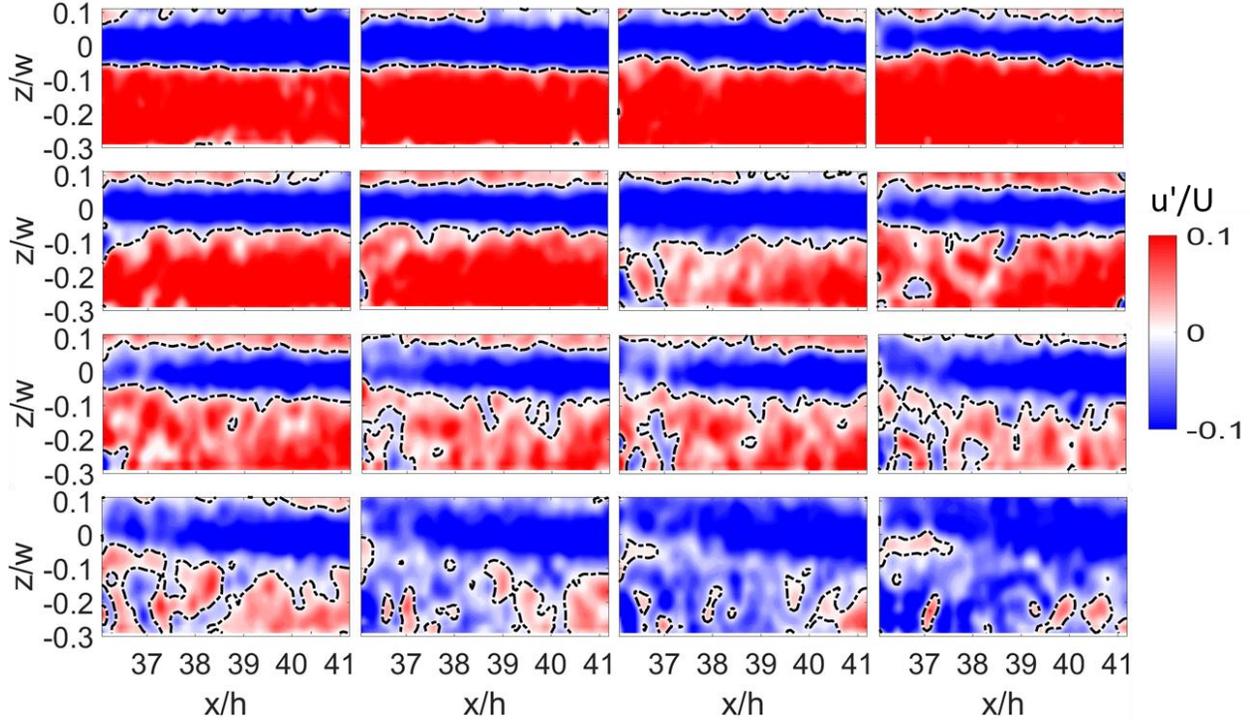

**Fig. S2.** **Next cycle to that is shown in Fig. 2 in ET at Wi=185.** The stream-wise streaks of the stream-wise velocity fluctuations ($u'$) in the frame moving with the mean velocity $u_m(z)$ normalized by U (spatio-temporal mean velocity from weighing balance) is presented with the scale on the right as a function of time at 16 moments of a single cycle with t* of 0.43, 0.49, 0.56, 0.62 (first row left to right), 0.74, 0.76, 0.8, 0.86 (second row left to right), 0.88, 0.89, 0.895, 0.9 (third row left to right), 0.91, 0.93, 0.95 and 0.97 (fourth row left to right), where t* is the time non-dimensionalised by the full cycle time of the breakdown (t* =t/$T_{cycle}$, $T_{cycle}$ = 350 msec). As in Fig. 2, the black dotted line of $u'$ marks the interface between the streaks moving in the opposite directions: with (red) and against (blue) the mean flow direction. The span-wise perturbations resulting from the KH-like pure elastic instability are developed temporally starting from almost unperturbed state.

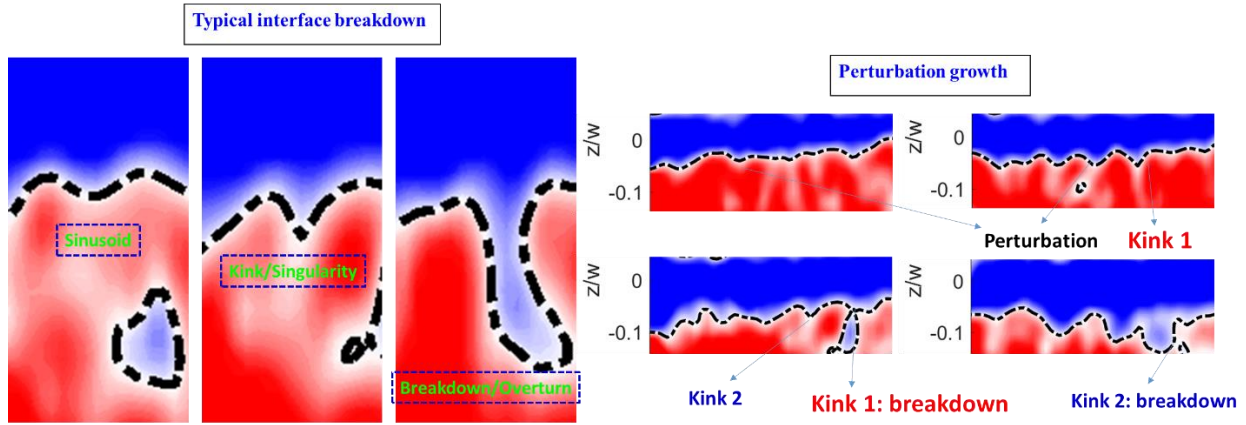

**Fig. S3.** Details of the perturbation growth, singularities, overturn and interface breakdown.

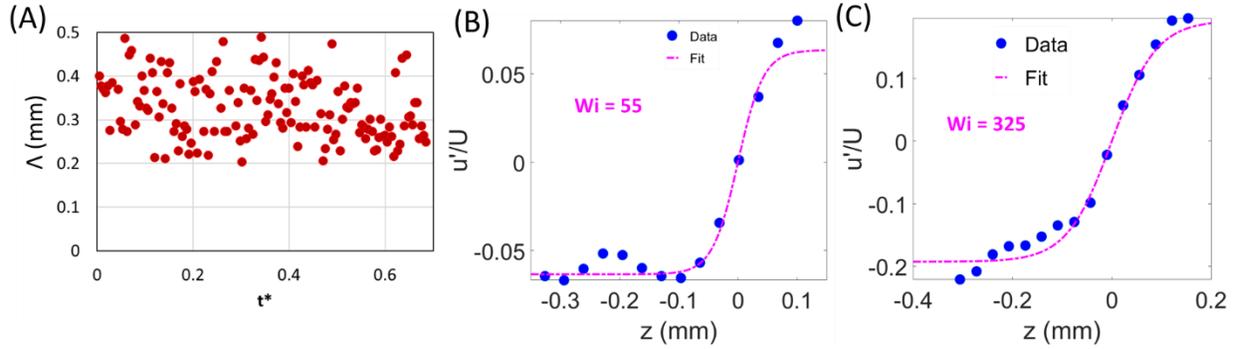

**Fig. S4.** (A) presents the temporal dependence of the perturbation wavelength Λ. It remains constant in time but with huge fluctuations, so Λ =0.35±0.15 mm. (B and C) show the data and the fits of velocity profiles on the interface at Wi of 55 and 325, respectively.

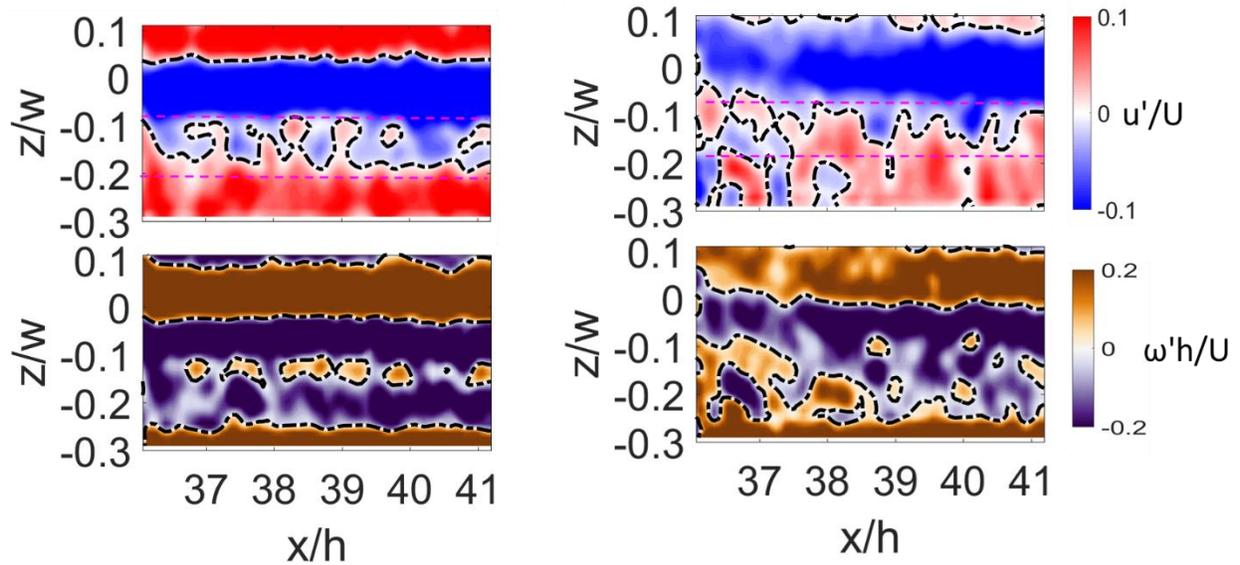

**Fig. S5. Vortex string, eddy size and mixing layer thickness.** Mixing layer, or a billow, thickness is marked by the magenta dashed lines. On the left hand side, the clear identified mixing layer is defined either by the size of the strong interface perturbations of the stream-wise velocity fluctuations $u'$ on the upper plot and by the width of the resulting vortex string in the vertical vorticity $\omega'$ presentation on the lower plot. On the right hand side, the identification of the mixing layer width is less definite but remains about in the range of 0.1 z/w. First column is for Run-1 at t* of 0.86 and second column is for Run-2 at t* of 0.9.

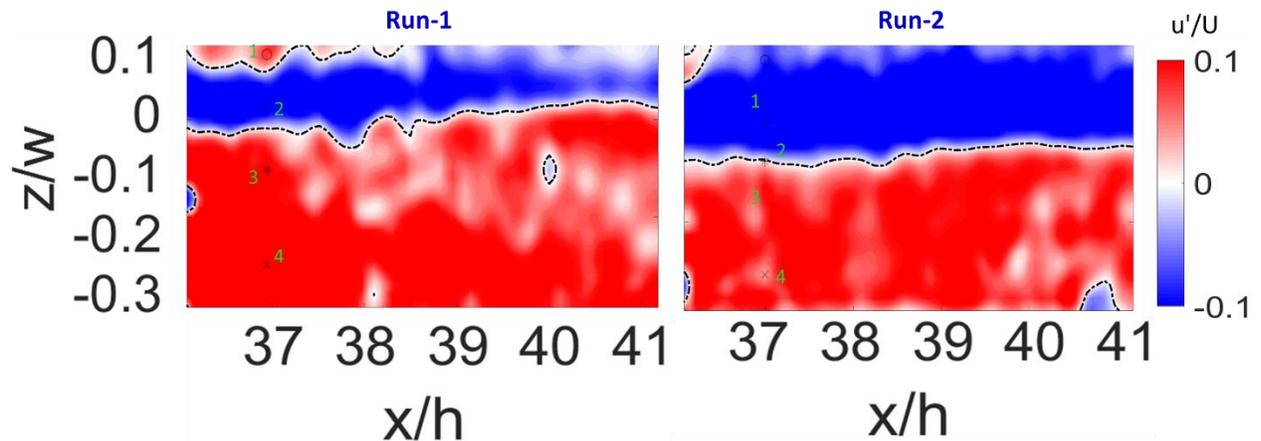

**Fig. S6.** **Marked location for the span-wise difference of the stream-wise velocity across the interface Δu=u₄-u₂ to study its time evolution during a cycle for two consequent cycles in ET at Wi=185.**

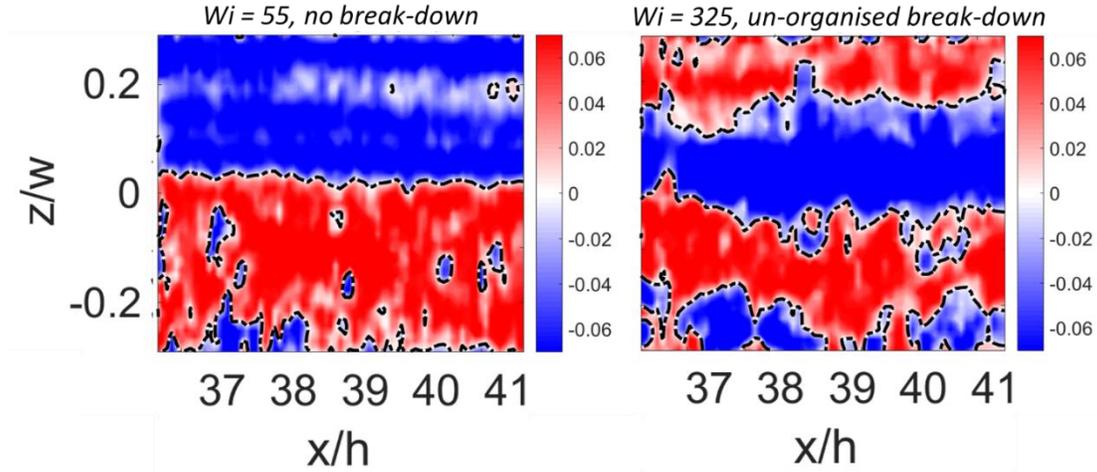

**Fig. S7.** **Perturbations of the interface and KH-like purely elastic instability in two flow regimes: transitional at Wi=55 and drag reduction (DR) at Wi=325.** At Wi=55 in the transitional regime, the perturbations remains low during a cycle, and the interface breakdown does not appear. Whereas at Wi=325 in the DR regime, the interface perturbations are huge, almost at the beginning of a cycle and a quicker breakdown to fully chaotic state takes place without any formation of the vortex string at the interface similar to Wi =185 case in Figure 2.

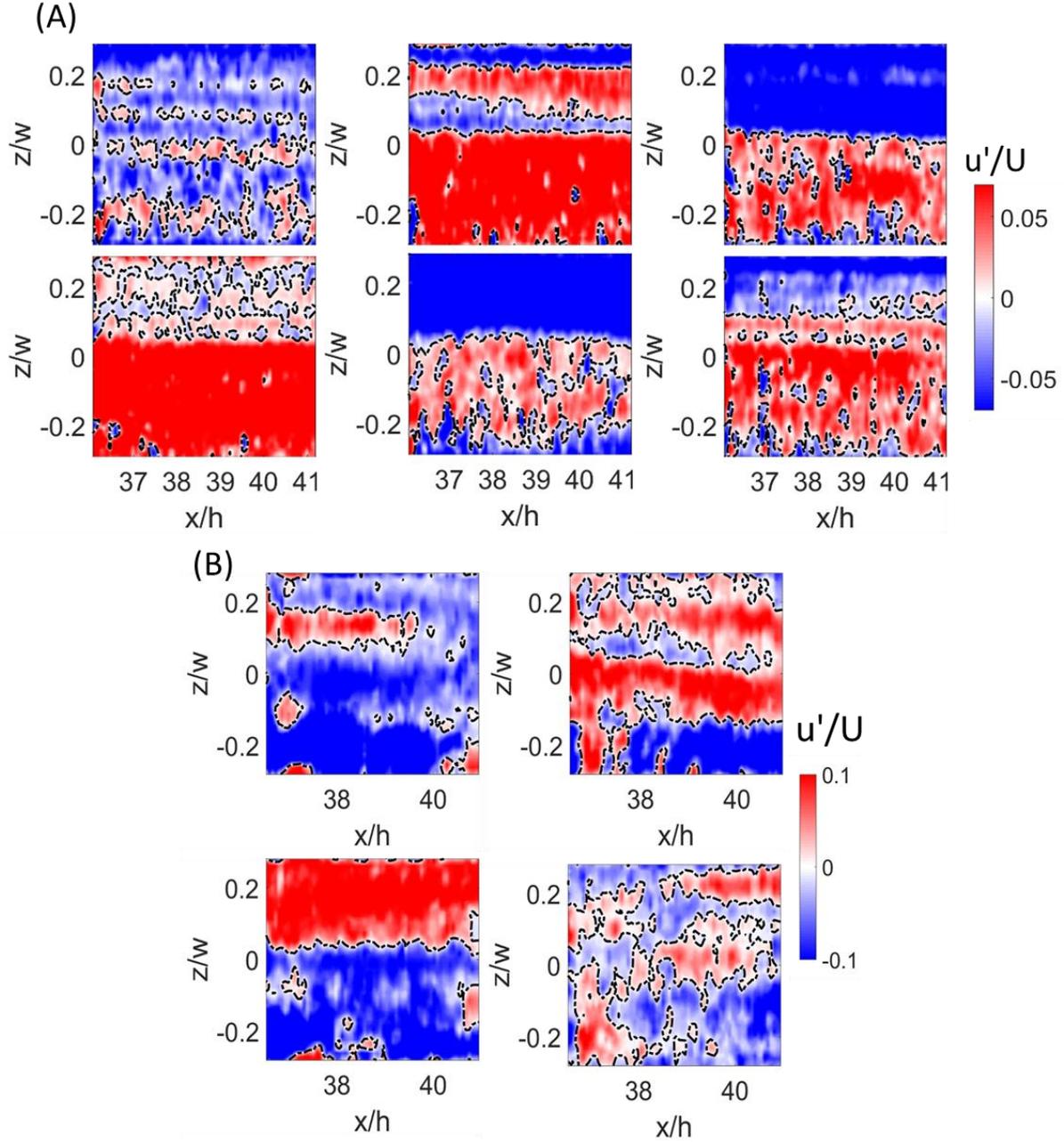

**Fig. S8.** **Examples of cycles of the elastic Kelvin-Helmholtz-like instability of the counter-propagating streaks interface in the transitional at Wi=55 (A) and drag reduction (DR) at Wi=325 (B) regimes.** Fluctuating stream-wise velocity $u'$ normalized by the mean stream-wise flow velocity U in xz-plane is presented with the scale shown on the right side of both data sets. The data are shown as a function of time at 6 moments at Wi=55 and 4 moments at Wi=325 in the corresponding moments in each cycle. At Wi=55, the data are taken at t* of 0.03, 0.2, 0.29, 0.37, 0.51 and 1, where t* is the time non-dimensionalised by the full cycle time of the breakdown (t* =t/$T_{cycle}$, $T_{cycle}$ = 2076 msec). At Wi=325, the data are taken at t* of 0.34, 0.43, 0.65 and 1 ($T_{cycle}$ = 279 msec).

**Movie S1.**

Temporal evolution of the stream-wise velocity streaks ($u'$) at Wi=185 during a cycle in ET regime. Evolution of the interface ($u' = 0$) can be observed from the perturbations to the stream-wise vortex roller and eventually to the complete breakdown of the streak towards the end of the cycle. Scale for the normalized stream-wide velocity ($u'/U$) are marked on the right side.

**Movie S2.**

Temporal evolution of the vertical vorticity ($\omega'$) at Wi=185 during a cycle in ET regime. "Nice" vortex string can be observed in the stream-wise vortex roller regime and then the eventual breakdown of the streak towards the end of the cycle. Scale for the normalized vertical vorticity ($\omega' h/U$) are marked on the right side.